# In Search of Excellence and Equity in Physics

Emanuela Barzi[*], *Fermi National Accelerator Laboratory, Batavia, IL 60510, USA, and Ohio State University, Columbus, OH 43210, USA*

S. James Gates, Jr., *Brown Theoretical Physics Center, Brown University, Providence, RI 02912, USA*

Roxanne Springer*, Duke University, Durham, NC 27708, USA*

[*] barzi@fnal.gov

Physicists often assert that most scientific work is judged on its merits and that opportunities in physics are equitably available to all aspirants. Equal opportunity is central to the concept of meritocracy. Opportunity and leadership should go to the people most qualified by performance, and not on the basis of arbitrary or irrelevant attributes. This principle is arguably most important for high-level leadership due to their outsized impact on the field. Aside from the ethical responsibility, research exists showing that traditionally underrepresented people draw new relations between ideas and concepts that lead to increased innovation [1]. Nevertheless, far worse underrepresentation persists than could be expected from a functioning meritocracy [2]. The expectation that our physics community is a meritocracy is thus challenged. If we want to change this, we need to change our behavior, i.e., practices. Practices of a community ought to be based on its stated underlying values. To begin, we should rigorously query assumptions that our organizations and the decisions made within them are equitable and seek good evidence that supports this assumption. As scientists, we need to learn from scientists who specialize in the study of human behavior.

Equitable processes and procedures, as opposed to hierarchical structures, do not occur naturally in human societies. Developing and maintaining fair procedures, and eliminating barriers to traditionally underrepresented people require focused effort, frequent measurement, and continuous correction. The risk along the way is that the concept of "Equity, Diversity, and Inclusion" (EDI) and associated activities become a meaningless checkbox [3], [4]. This risk is understandable from a sociological point of view; those in power are less aware than others in their organization of the harmful consequences of inequity, as well as being poor judges of their own competence as leaders. Thus, organizations tend to respond to demands for change by offering the appearance of change, without monitoring mechanisms or measurable impacts.

The importance of equity in the workplace is discussed by E. Kevin Kelloway, the Canada Research Chair in Occupational Health Psychology at St. Mary's University, who says "Injustice is a particularly toxic stressor because it strikes at the core of who we are … When you treat me unfairly, you attack my dignity as a person — essentially saying that I don't deserve fair treatment or to be treated the same as others." [5] Increased stress can impact human health by compromising the immune system. Stress effects on the body are described in a study by the

American Psychological Association [6]. Jeffrey Pfeffer, Thomas D. Dee II Professor of Organizational Behavior at Stanford University, writes, ``Job stress costs US employers more than $300 billion annually and may cause 120,000 excess deaths each year." [7] The effect on science generally from bullying and/or unequal opportunity is presented in a report of the National Academies [2].

To address this, the physics community should implement best practices on how to unite excellence and equity from other very competitive fields. The private sector has already done much work in mitigating the financial and reputational damages that followed the public exposures of misconduct by the #MeToo and similar movements. In business, both the financial bottom line and the swiftness of information flowing to potential customers forces nimble adjustments. When inequity threatens the success of a business, the business changes its practices or fails. Therefore, corporations have been increasing research funding of Industrial-Organizational Psychology (I/O Psychology), the aim of which is to match behavior, i.e., best practices, to the values of an organization.

The most prevalent forms of discrimination include harassment (verbal and nonverbal behaviors that convey hostility, objectification, exclusion, or second-class status) and retaliation. In discrimination, the most potent predictor is the degree to which misconduct is perceived to be tolerated in the organization [2]. Institutions can reduce harassment and retaliation by making systemic changes to demonstrate that misconduct will not be tolerated and reports of misconduct will be taken seriously. Rigid bureaucratic/severe hierarchical structures and concentrated power centers often are correlated with fostering and sustaining harassment and promoting retaliation [8]. Organizations that enable a climate of aggression and bullying are more likely to have managers who abuse power, and who are more likely to ruminate over perceived offenses, ultimately seeking retaliation [8].

Efforts to create an equitable, diverse, and inclusive discipline are currently being undertaken at the American Physical Society [9], [10], [11], [12]. But how do we implement the changes needed in the physics community at large to improve our excellence and equity? An important starting point is how we choose our leaders for high-responsibility positions in academia, national labs, managing organizations, funding agencies, and industry. At the moment, many in the community perceive that the choice of leaders is infused with a lack of meritocracy and too often driven by cronyism.

Our community needs leaders who exhibit the highest integrity, who inspire others, who own their mistakes, who continue to learn and grow in the face of new evidence and new circumstances, who value equity and excellence over the appearance of it, who understand that an organization's culture is dictated by the experience of its most vulnerable members, and who will put facts and fair play above politics and convenience.

To identify such a leader, searches can follow best practices developed by I/O Psychology. The National Science Foundation has also been funding studies in this area [13], [14]. An ethical hiring process for leaders who are excellent and will promote equity should include:

1. Identifying the characteristics needed for a successful performance in the leadership role. In I/O Psychology this is part of a quantitative process called "job analysis."
2. Advertising open leadership positions broadly, including among professional societies who reach traditionally underrepresented people, for example, the National Society of Black Physicists, the National Society of Hispanic Physics, the National Technical Association, and the Society for the Advancement of Chicanos/Hispanics and Native Americans in Science.
3. Creating search committees that consist of a diverse representation of stakeholders, measured by position, location, race, gender, etc.; inviting representation from advocacy groups within the organization.
4. Training search committees to identify their own biases and that of others (for example, when reading letters of recommendation) and to mitigate them. The training should address not just legal guardrails but practices that have been shown to improve equitable hiring.
5. Agreeing on a rubric before examining applications. This rubric should be based on the characteristics identified in item (1), prioritize the characteristics, and create a normalized scoring procedure.
6. Building the selection process to include only steps that add value and measurement to the process; measuring the critical skills and abilities from item (1) multiple times throughout the process; informing everyone, and particularly the selection committee and the stakeholders, about the process.
7. Using reliable assessment tools [15] so as to add value in the form of predicting future behavior since assessments consistently outperform and out-predict all other aspects of the hiring process.
8. Interviewing candidates using a structured and consistent process across applicants, including personality evaluations, integrity and reliability tests, work simulations, i.e. both situational and behavioral interview questions. Behavior research shows that unstructured interviews are two to three times less reliable than a structured process [16].
9. Inviting candidates to submit a short statement that does not contain their name or current institution, but instead lists how they qualify for the leadership position, why they want it, and what they wish to accomplish in the leadership position, including efforts to create and sustain a fair and equitable work environment.
10. Providing each candidate with a case study that involves an EDI related problem that they have not seen in advance, where they are invited to discuss how they would handle the problem were they to be hired into the leadership position.
11. Considering the appropriateness of confidentiality of the leadership candidate names. When stakeholders are invited to provide feedback, the search committee obtains valuable information that reduces the risk of unexpected and potentially embarrassing information arriving after a new leader is announced.

When a transparent and equitable search that follows best practices is not performed, it sets the tone for the lack of trust the stakeholders will have in leaders of the institution. Further, behavioral change cannot be accomplished without the advocacy of leaders; institutions like physics departments, national labs, etc. cannot be healthier than the organizations that manage them. Likewise, it is also in the interest of funding agencies, as stewards of taxpayer's money, to require ethical practices and monitor behavior for the most cost-effective use of their money.

Examples of what happens if equity is not prioritized are given in [17], [18], [19]. We can do better. If we expect to keep the best and brightest in our field, then we must do better. Discounting deserving, productive scientists from the scientific enterprise, be it for inequity, exclusion of minorities or even illegal discriminatory behavior, undermines the integrity of science and affects it at its core now and in the future.

**Acknowledgments**
The authors are grateful to Samuel Bader, Jonathan Bagger, William Barletta, Charles Bennett, Banafsheh Ghassemi, Matthew Hannah, Nida Hussain, Joseph Niemela, and Jeannette Russo for their valuable feedback.

[19] https://www.cnn.com/2022/02/07/politics/eric-lander-white-house-investigation/index.html

Also in the APS CSWP & COM Gazette, Spring 2022 issue, at
https://www.aps.org/programs/women/reports/gazette/upload/Spring22.pdf